\documentclass[final]{aipproc}
\layoutstyle{6x9}

\begin{document}
\title{A narrow "peanut" pentaquark}

\author{Dmitri Melikhov}
{address={Nuclear Physics Institute, 
  Moscow State University, 119992, Moscow, Russia}}
\begin{abstract}
We analyse the decay $\Theta_s(1/2^+)\to NK$ in a non-relativistic Fock space description
using three and five constituent quarks for the nucleon and the pentaquark, respectively. 
Following Jaffe and Wilczek \cite{jw}, we assume that  
quark-quark correlations in spin-zero state play an important role 
for the pentaquark internal structure. 
Within this scenario, a strong dynamical suppression of the decay width is shown to be 
possible only if the pentaquark has an asymmetric "peanut" structure with the strange
antiquark in the center and the two extended composite diquarks
rotating around. In this case a decay width 
of $\simeq$ 1 MeV may be a natural possibility. 
\end{abstract}
\maketitle

The existence of pentaquarks is not yet undoubtedly established. 
But if these particles exist, the exotic members of the pentaquark
multiplet must have a very small decay width of order 1 MeV or even 
lower. For the possible origin of the small pentaquark width 
many qualitative suggestions have been put forward. 
In a scenarios proposed by Jaffe and Wilczek \cite{jw} the positive-parity spin-1/2 pentaquark 
consists of an antiquark and two scalar diquarks in a relative 
$P$-wave state. 
In this talk I present the results of a fully dynamical quark-model calculation 
of the pentaquark width 
done together with B.Stech and S.Simula \cite{mss} using a non-relativistic Fock 
space representation for the $J^P=\frac12^+$ pentaquark in the Jaffe-Wilczek scenario. 

The decay amplitude $T(\Theta\to KN)$ is related 
to the matrix element 
\begin{eqnarray}
\label{2.7}
\langle N(p')|\bar s\gamma_\mu \gamma_5 d|\Theta(p)\rangle
&=& 
g_A(q^2) \bar u_N(p')\gamma_\mu\gamma_5 u_\Theta(p)
+
g_P(q^2) q_\mu \bar u_N(p')\gamma_5 u_\Theta(p) \nonumber\\ 
&&+
g_T(q^2) \bar u_N(p')\sigma_{\mu\nu}q^\nu \gamma_5 u_\Theta(p), \qquad \qquad q=p-p'.
\nonumber
\end{eqnarray}
Here the form factors $g_i$ contain poles at $q^2>0$ due to strange meson resonances 
with the appropriate quantum numbers.  
The residue of the pole in $g_P$ at $q^2=M_K^2$ is related to the amplitude of interest 
$T(\Theta\to NK)$: for $q^2\to M_K^2$   
$$
(M_K^2-q^2)g_P(q^2) \bar u_N(p')\gamma_5 u_\Theta(p)\to f_K {T(\Theta\to NK)},
$$  
where $f_K=160$ MeV is the kaon decay constant. The form factor $g_A$ 
contains the pole at $q^2=(K^*_A)^2$, but at $q^2=M_K^2$ it is a
regular function. Making use of the relationship between the form factors 
$g_A$ and $g_P$ emerging in the limit of spontaneously broken chiral symmetry \cite{mss} gives  
$$
T(\Theta\to NK)=\frac{M_\Theta+M_N}{f_K}g_A(M_K^2)\cdot \bar u_N(p')i\gamma_5 u_\Theta(p)   
$$
and 
$$
\Gamma(\Theta)=\Gamma(\Theta\to K^+n)+\Gamma(\Theta\to K^0p) 
\simeq\frac{1}{\pi}\frac{|\vec q|^3}{f_K^2}g_A^2(M_K^2). 
$$
For $M_\Theta=1540$ MeV one finds $|\vec q|=270$ MeV and 
$\Gamma(\Theta)= 240\; g_A^2$ MeV. 
For transitions between hadrons of the same quark structure $g_A\simeq 1$ 
(e.g. for the nucleon $g_A\simeq 1.23$). So for a normal resonance   
one would expect $\Gamma(\Theta)\simeq 200$ MeV.  
To obtain a width of $\le$ 10 MeV one needs a strongly suppressed value $g_A\le 0.2$. 

In \cite{mss} we calculated the amplitude  
$\langle N|\bar s \gamma_\mu \gamma_5 d|\Theta\rangle$ 
and the form factor $g_A(q^2)$ using a non-relativistic equal-time Fock space representation.  

\noindent{\bf The nucleon} in this framework is described by its coordinate wave function 
depending on the relative coordinates 
$\vec\rho_N=\vec r_2-\vec r_3$ and $ 
\vec\lambda_N=\frac1{2}(\vec r_2+\vec r_3)-\vec r_1$, for which we take the Gaussian 
function  
$$
\Psi_N(r_1|r_2,r_3)\sim\exp\left(-\frac{1}{2\alpha^2_{\rho N}}\vec\rho_N^2 
-\frac{2}{3\alpha^2_{\lambda N}}\vec\lambda_N^2\right). 
$$
{\bf The pentaquark} coordinate wave function depends on the relative coordinates 
$\vec{r}_{23}=\vec{r}_2-\vec{r}_3$,  
$\vec{R}_{23}=\frac{1}{2}(\vec{r}_2+\vec{r}_3)$,
$\vec{r}_{45}=\vec{r}_4-\vec{r}_5$, 
$\vec{R}_{45}=\frac{1}{2}(\vec{r}_4+\vec{r}_5)$,  
$\vec{\rho}_\Theta=\vec{R}_{23}-\vec{R}_{45}$, 
$\vec\lambda_\Theta=\frac{1}{2}(\vec{R}_{23}+\vec{R}_{45})-\vec{r}_1$, 
where $\vec r_1$ is the position of the strange particle, $\vec R_{23}$ and $\vec R_{45}$ 
are the positions of the two diquarks. As required by the quark-diquark scenario, 
the pentaquark coordinate wave function factorizes into the diquark wave functions and the wave function 
of the three-particle quark-diquar-diquark system, for which we take again Gaussian parameterizations 
\begin{eqnarray}
\label{pqwf}
\Psi_\Theta(r_1|r_2,r_3|r_4,r_5)\sim 
\exp\left(-\frac{1}{2\alpha^2_{\rho \Theta}}\vec{\rho}_\Theta^2 
-\frac{2}{3\alpha^2_{\lambda \Theta}}\vec{\lambda}_\Theta^2\right)
\exp\left(-\frac{\vec{r}^2_{23}}{2\alpha_D^2}\right)
\exp\left(-\frac{\vec{r}^2_{45}}{2\alpha_D^2}\right).\nonumber    
\end{eqnarray}
{\bf The form factor} $g_A$ can be expressed through the following vector overlap amplitude 
\begin{eqnarray}
&&\frac{24}{\sqrt{3}}
\int d\vec r_2 d\vec r_4 d\vec r_5 
\exp\left(i\vec q\,\frac{\vec r_2+\vec r_4+\vec r_5}{3}\right)\nonumber \vec \rho_\Theta\;
\Psi_\Theta(r_s|r_2,r_d|r_4,r_5)\\
&&\qquad\qquad\qquad\qquad\qquad\qquad\qquad\qquad\qquad\times
\left\{2\Psi_N(r_2|r_4,r_5)+\Psi_N(r_4|r_2,r_5)\right\}, \qquad 
\nonumber\\
&&
\vec \rho_\Theta=\frac12(\vec r_2+\vec r_d-\vec r_4-\vec r_5). 
\nonumber 
\end{eqnarray}
Details of this calculation can be found in our paper \cite{mss}. 

\noindent{\bf Numerical estimates}. 
We present now numerical results for the pentaquark width. 
Two assumptions reduce the number of parameters:   

\noindent 
1. The structure of the diquark in the nucleon and in the pentaquark coincide, i.e.   
the size-parameter $\alpha_D$ of the diquark wave function $\Phi_D$ is equal to the parameter 
$\alpha_{\rho N}$ of the nucleon wave function, $\alpha_D=\alpha_{\rho N}$. 

\noindent 
2. The parameters of the nucleon wave function are chosen such that the experimental 
nucleon electromagnetic form factor is reproduced for small momentum transfers,   
${\alpha^2_{\lambda N}}/{16}+{\alpha^2_{\rho N}}/{48} = 1/{M_\rho^2}$.
We first take a symmetric wave function $\alpha_{\lambda N}=\alpha_{\rho N}=0.9$ fm.
The diquark size parameter is then 
$\alpha_D=\alpha_{\rho N}=0.9$ fm. Now only the two free parameters of the pentaquark wave function 
 $\alpha_{\rho\Theta}$ and $\alpha_{\lambda\Theta}$ remain to be fixed. 
Recall that $\alpha_{\rho\Theta}$ determines the average distance between the 
two extended diquarks, and $\alpha_{\lambda\Theta}$ determines the average distance between 
the $s$-antiquark and the center-of-mass of the two diquarks. 
Little is known about the details of the pentaquark structure. Therefore 
we allow  the parameters 
$\alpha_{\lambda\Theta}$ and $\alpha_{\rho\Theta}$ to vary in a broad range  
$
0.6\; fm<\alpha_{\lambda\Theta},\;\alpha_{\rho\Theta} < 1.6\; fm  
$
and study the dependence of $g_A$ and the width on these parameters. 

Fig. 1(a) shows $\Gamma(\Theta)$ vs the pentaquark size parameters
$\alpha_{\lambda\Theta}$ and $\alpha_{\rho\Theta}$. 
If both parameters are $\simeq 1$ fm, then $g_A\simeq 0.8$ and the width is $150$ MeV. 
{No suppression due to a possible mismatch of color and flavour quantum numbers in the 
initial and final states takes place}. However, 
a strong dynamical suppression occurs if the structure of the
pentaquark is asymmetric:  
For instance, for $\alpha_{\lambda\Theta}=0.6 \; fm, \alpha_{\rho\Theta}=1.4 \; fm$, 
we get $g_A=0.05$ and $\Gamma(\Theta)=1$ MeV. 
  
Fig. 1(b) presents $\Gamma(\Theta)$ vs the diquark size $\alpha_D$ 
for fixed values of the pentaquark size-parameters
$\alpha_{\rho\Theta}=\alpha_{\lambda\Theta}=1$ fm. 
A sizeable reduction of the pentaquark width occurs only for a very
small diquark size which corresponds to implausibly large deviations
from a symmetric nucleon wave function. Such compact diquarks are 
not supported by a successful description of the nucleon properties
with a symmetric wave function. 
\begin{figure}[h]
\begin{tabular}{cc}
\includegraphics[totalheight=6.5cm]{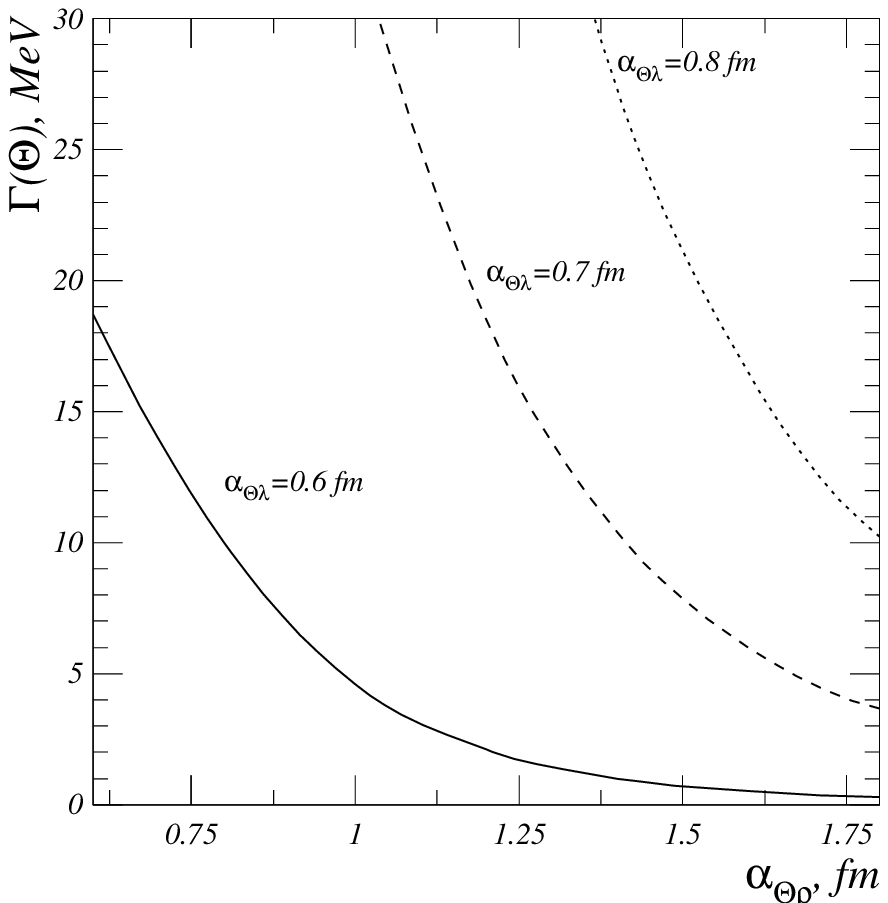}&
\includegraphics[totalheight=6.5cm]{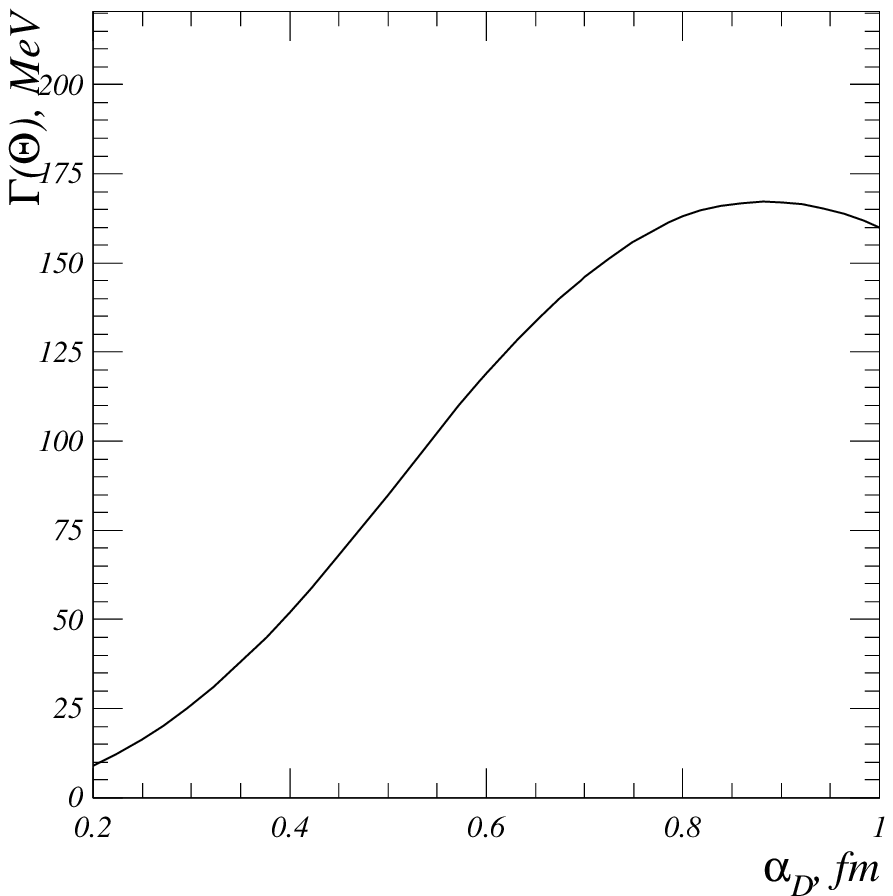}
\end{tabular}
\caption{\label{fig:1}
{\it Left (a)}: $\Gamma(\Theta)$ vs 
the pentaquark size parameters $\alpha_{\rho\Theta}$ 
and $\alpha_{\lambda\Theta}$. 
{\it Right (b)}: $\Gamma(\Theta)$ vs the diquark size parameter $\alpha_D$ for a symmetric 
pentaquark $\alpha_{\Theta\rho}=\alpha_{\lambda\Theta}=1$ fm.} 
\end{figure}

Summing up, the pentaquark decay width $\Gamma(\Theta)$ is found to depend strongly on the pentaquark configuration: 
when all size-parameters of the pentaquark wave function are close to 1 fm, one obtains a width of 
about $150$ MeV, i.e. a typical hadronic value. 
{\it The color-flavour structure of the pentaquark causes no suppression of the width}.\footnote{ 
For a discussion of the pentaquark width in the chiral limit we refer to \cite{ms}.}

A strong dynamical suppression of the amplitude occurs for a "peanut"-shaped  
pentaquark, i.e. when it has an asymmetric structure with 
$\alpha_{\lambda\Theta}\ll \alpha_{\rho\Theta}$. 
For instance, $\alpha_{\lambda\Theta}=0.6$ fm and $\alpha_{\rho\Theta }=1.4$ fm 
brings the width down to 1 MeV. 

We therefore conclude that {\it if the pentaquark can be described as a five-quark system, 
in which two composite spin-zero diquarks are in the relative $P$-wave state, 
the small width requires a rather asymmetric "peanut" structure with two 
extended diquarks rotating about the strange antiquark localized near the center}. 

\vspace{-.3cm}
\theacknowledgments
I would like to thank my friends Berthold Stech and Silvano Simula for the most 
enjoyable collaboration, and Nora Brambilla for the invitation to participate 
in this exciting Conference. 

\end{document}